\documentclass[journal]{IEEEtran}

\usepackage{multicol}
\usepackage{multirow}

\usepackage{cite}

\ifCLASSINFOpdf

\else

\fi

\usepackage[cmex10]{amsmath}
\usepackage[cmex10]{amsmath}
\usepackage{amssymb}

\usepackage{array}
\usepackage{mdwmath}
\usepackage{mdwtab}

\newcommand{\ie}{\emph{i.e. }}
\newcommand{\etal}{\textit{et al. }}

\usepackage[dvips]{graphicx}
\graphicspath{{eps/}}
\DeclareGraphicsExtensions{.eps}

\IEEEoverridecommandlockouts

\newtheorem{definition}{Definition}
\newtheorem{example}{Example}

\begin{document}
%
\title{Design of Convergence-Optimized Non-binary LDPC Codes over Binary Erasure Channel}

\author{{Yang Yu, Wen~Chen~\IEEEmembership{Senior Member,~IEEE}, and Lili Wei~\IEEEmembership{Member,~IEEE}}
\thanks{Manuscript received April 20, 2012; accepted April 23, 2012. The associate editor coordinating the
review of this paper and approving it for publication was Giulio
Colavolpe.}
\thanks{The authors are with Department of Electronic Engineering, Shanghai Jiao Tong University,
China; W. Chen is also With SKL for ISN, Xidian University, China,
e-mail: \{yuyang83, wenchen, liliwei\}@sjtu.edu.cn.}
\thanks{
This work is supported by national 973 project \#2012CB316106, by
NSF China \#60972031 and \#61161130529, by national 973 project
\#2009CB824904, by national key laboratory project \#ISN11-01.}}

\markboth{IEEE Wireless Communications Letters. VOL. X. NO. X. XXXXXX 2012}%
{\MakeLowercase{Design of Convergence-Optimized Non-binary LDPC Codes over Binary Erasure Channel}}

\maketitle

\begin{abstract}
In this letter, we present a hybrid iterative decoder for non-binary
low density parity check (LDPC) codes over binary erasure channel (BEC),
based on which the recursion of the erasure probability is derived to
design non-binary LDPC codes with convergence-optimized degree distributions.
The resulting one-step decoding tree is cycle-free and achieves lower decoding complexity.
Experimental studies show that the proposed convergence-optimization algorithm
accelerates the convergence process by $33\%$.

\end{abstract}


%
\IEEEpeerreviewmaketitle

\section{Introduction}
The erasure channel describes a common phenomenon in some communication and storage systems:
a symbol transmitted over this channel is either received or erased with certain probability.
Since there is no method to guarantee the accuracy of the raw data transmitted over this channel,
one of the most important methods is to use forward error-correcting (FEC) codes, among which LDPC codes have been shown of great potential in approaching the theoretical error correction limits\cite{sv08DensityEvolution,ge09NLDPC}.
More importantly,
the threshold-predicting procedure for the coded bits transmitted over the erasure channel
can be easily exploited to predict the (approximated) performance-threshold for other channels as well,
such as AWGN channels, binary symmetric channels \etal \cite{Ashi04ExitErasure}.


Investigation over finite field, \ie ${\mathbb F}_{q}$, $q=2^p$,
shows that the $q$-ary LDPC codes have better performance than the binary LDPC codes
for not very long block length,
and irregular LDPC codes can outperform the regular LDPC codes.
In this letter, we present a hybrid iterative decoder
for $q$-ary LDPC codes over BEC in section \ref{sec_iterative_decoder} by using their binary images.
Then the recursion of erasure probability of this decoder
is derived in section \ref{sec_convergence_optimize}.
Further, we give an irregular optimization algorithm to design
$q$-ary LDPC codes with convergence-optimized degree distributions.

The advantage of our proposed algorithms is obvious:
\emph{(i)} when a symbol is represented by its binary image,
  the bits within a binary image are not independent regarding the decoding procedure,
  \ie there exist cycles within
  the binary image.
  As a result, this inner dependence introduces more unrecoverable bits.
  \cite{Savin09Bec} solves this problem by adding
  a \emph{simplex} constraint to the check nodes.
  Although they use density evolution to find threshold-optimized codes,
  recursion of the erasure probability for the original bits
  under this constraint is still implicit.
  The proposed hybrid iterative decoder removes the dependence by putting an additional
  reverse operation while decoding each symbol. Each resulting
  binary image is cycle-free
  whose convergence behavior can be explicitly characterized by a recursion of the erasure probability.
\emph{(ii)} Based on this recursion, an optimization algorithm is proposed
  to design $q$-ary LDPC codes whose decoder converges faster in the lower channel erasure probability
  regime than the threshold-optimized $q$-ary LDPC codes.
\emph{(iii)} 
  Computational complexity of the decoding algorithm for $q$-ary LDPC codes over BEC proposed in
  \cite{Savin09Bec} is dominated by $O(q)$ for each check-sum operation.
  In this letter, we reduce the computational complexity to $O(\log^2 q)$ which
  is smaller than $O(q)$ for large $q$.

\section{Iterative Hybrid Decoder and Code Design}\label{sec_iterative_decoder}
\subsection{Equivalent Binary Codes Construction}\label{sec_binary_codes}
We consider a non-binary LDPC code $\mathcal{C}$ with parity check matrix $\mathbf H$
defined over finite field $\mathbb F_q$.
The entries of the associated parity check matrix $\mathbf H$ are also called \emph{labels} along the
corresponding edges in the Tanner graph.
Then by assuming $\mathbb F_q$ be endowed with a vector space structure over $\mathbb F_2$
and letting matrix $\mathbf A$ be a \emph{canonical cyclic generator} of $\mathbb F_q$,
\ie $\mathbb F_q \cong \{0,\mathbf A^i,0\leqslant i\leqslant q-2\}$ \cite{lidl2002,xiao07UniDecLdpc},
we are ready to have our definition of the equivalent binary LDPC code $\mathcal{\bar C}$.
\begin{definition}
  Let $\mathcal C$ be a non-binary LDPC code with
  parity check matrix $\mathbf H=(h_{m,n})_{M\times N}$.
  Each codeword $\mathbf x = (x_1,x_2,\ldots,x_N)$ in $\mathcal C$ can be
  represented by its binary form $\bar{\mathbf x} = (\bar{\mathbf x}_1^T,\bar{\mathbf x}_2^T,\ldots,\bar{\mathbf x}_N^T)$, where $\bar{\mathbf x}_i^T$ is (row) vector representation
  of $x_i$.
  Then the binary LDPC code $\bar{\mathcal C}$ associated with the non-binary LDPC code $\mathcal C$ is defined by
  \begin{eqnarray}
    \nonumber
    \bar{\mathcal C} &=& \ker(\bar{\mathbf H}) \subset \mathbb F_2^{Np}\\
    \nonumber
    &=& \left\{(\bar{\mathbf x}_1^T,\bar{\mathbf x}_2^T,\ldots,\bar{\mathbf x}_N^T)|
    \sum_{n=1}^{N} \mathbf A_{m,n}\bar{\mathbf x}_n=0,\forall m = 1,\ldots, M\right\},
  \end{eqnarray}
  where $\bar{\mathbf H}$ is parity check matrix resulting from replacing each entry $h_{m,n}$ in
  $\mathbf H$ by its matrix representation $\mathbf A_{m,n}$, \ie \emph{matrix} label.
\end{definition}

\subsection{Decoding Algorithm}
Tanner graph $\bar{\mathcal G}$ of $\bar{\mathcal C}$ is equivalent to
the Tanner graph $\mathcal G$ of $\mathcal C$
in the sense that a $q$-ary symbol in $\mathcal G$ can be viewed as
a binary vector in $\bar{\mathcal G}$. The relationship
between vectors is equivalent to the relationship between the $q$-ary symbols
regarding the parity check functions.
The inverse operation utilizes this relationship, which makes the hybrid iterative
decoding algorithm in general equivalent to the $q$-ary maximum-likelihood decoding algorithm described
in \cite{Savin09Bec}.
However, characterization of the decoding over $\bar{\mathcal G}$
is much different from the decoding over $\mathcal G$.
This is because that
\emph{(i)} After transmitting the codeword from $\mathcal{\bar{C}}$ over BEC,
the received bits are not converted to their associated $q$-ary symbols.
Decoding over $\bar{\mathcal G}$ is performed on bits.
A non-binary symbol is recovered iff bits in its binary image are either recovered or not lost;
\emph{(ii)} Constituent bits in a binary image are decoded dependently, \ie
if a variable node $n$ is connected to a check node $m$ in $\mathcal G$,
then there are cycles between the bit-vector node $\bar{\mathbf n}$ and the check-vector node
$\bar{\mathbf m}$ under the equivalent representation in $\bar{\mathcal G}$
as shown in example~\ref{eg_inverse_operation};
\emph{(iii)} Randomly generated labels make the weights of rows and columns of the equivalent matrix representations
non-constants, \ie degree distribution of $\bar{\mathcal G}$ are not a straightforward
extension of the degree distribution of $\mathcal G$.
Regarding the problems, we design more tractable decoding algorithm in this letter.

In \cite{Savin09Bec}, the authors put a \emph{simplex} constraint on check nodes
which removes the dependence of bits in a binary image.
But the degree distribution of the original bits is still implicit. We
solve this problem by adding a reverse operation while decoding a bit-vector node that results in
a hybrid iterative decoding process over $\bar{\mathcal G}$.
We give the details in the following example.

\begin{figure}[!hbtp]
\begin{center}
\includegraphics[width=0.4\textwidth]{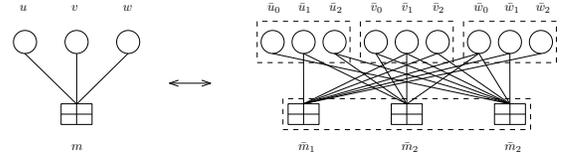}
\end{center}
\vspace*{-.3cm}
\caption{Equivalent binary check graph.}
\label{fig_Equivalent_binary_check}\vspace*{-.3cm}
\end{figure}

\begin{figure}[!hbtp]
\begin{center}
\includegraphics[width=0.4\textwidth]{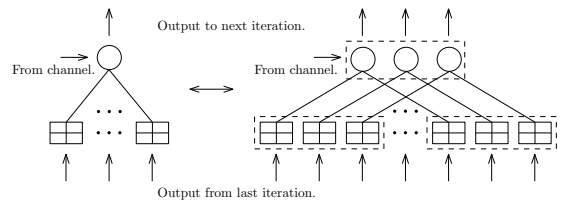}
\end{center}
\vspace*{-.3cm}
\caption{Equivalent binary one-step decoding tree of $\hat{\mathcal G}$.}
\label{fig_Equivalent_decoding_tree}\vspace*{-.3cm}
\end{figure}

\begin{example}\label{eg_inverse_operation}
  We consider the $q$-ary LDPC code $\mathcal C$ over $\mathbb F_8$.
  Codeword $\bar{\mathbf x}$ taken from $\bar{\mathcal C}$ is transmitted over BEC.
  We assume that every bits of $\bar{\mathbf x}$ are erased with the same channel erasure probability.
  Let $\mathcal H(m)=\{n|h_{m,n}\neq 0\}$ be the set of variable nodes participating in check node $m$,
  and $\mathcal N(n)=\{m|h_{m,n}\neq 0\}$ be the set of check nodes connected to variable node $n$.
  Considering a variable node set $\mathcal H(m)=\{u,v,w\}$ regarding the check node $m$
  of degree-$3$. Then the parity check equation is
  \begin{equation}\label{eqn_eg_nonbinary}
    \alpha^7 u + \alpha^5 v + \alpha^4 w = 0,
  \end{equation}
  where $\alpha$ is the \emph{primitive element} of $\mathbb F_8$\cite{lidl2002}.
  Considering the equivalent binary code $\bar{\mathcal C}$, Eq.~\eqref{eqn_eg_nonbinary}
  becomes
\begin{equation}\label{eqn_eg_binary}
    \mathbf A^{2}\underbrace{\left(\begin{array}{c}
      \bar{u}_0\\
      \bar{u}_1\\
      \bar{u}_2\\
    \end{array}\right)}_{\bar{\mathbf u}}
    +\mathbf A^{4}\underbrace{\left(\begin{array}{c}
      \bar{v}_0\\
      \bar{v}_1\\
      \bar{v}_2\\
    \end{array}\right)}_{\bar{\mathbf v}}
    +\mathbf A^{5}\underbrace{\left(\begin{array}{c}
      \bar{w}_0\\
      \bar{w}_1\\
      \bar{w}_2\\
    \end{array}\right)}_{\bar{\mathbf w}}=0.
  \end{equation}

  The equivalent binary graph for the non-binary parity check equation
  is given in Fig.~\ref{fig_Equivalent_binary_check}. If the
  decoder attempts to decode $\bar{\mathbf w}$ in Eq.~\eqref{eqn_eg_binary}, we first
  multiply the inverse of its associated matrix label, \ie $\mathbf A^{-5}$, to each
  additive items in Eq.~\eqref{eqn_eg_binary}. Then we move $\bar{\mathbf w}$ to
  the other side of the equal sign. Since $\bar{\mathbf w}$ is a binary vector,
  the inverse of $\bar{\mathbf w}$ is itself. As a result, we have
  \begin{eqnarray}\label{eqn_inverse_operation}
    \bar{\mathbf w}&=&\mathbf A^{2-5}\bar{\mathbf u}+\mathbf A^{4-5}\bar{\mathbf v},
  \end{eqnarray}
\end{example}
where the addition at the exponent of the companion matrix is performed over the ring of integers mod $(q-1)$.
Eq.~\eqref{eqn_inverse_operation} is actually a equation set representing the parity-check relationship
within the bit-vectors. Each erased bit in $\bar{\mathbf w}$ can be recovered iff
the other bits participating in its parity-check equation are not lost.
Note that Eq.~\eqref{eqn_inverse_operation} is different from Eq.~\eqref{eqn_eg_binary}
in the sense that each constituent bit in $\bar{\mathbf w}$ can be decoded independently
in Eq.~\eqref{eqn_inverse_operation}.
In other words, the inverse operation added to each tentative decoding
transforms the equivalent one-step decoding tree into a more tractable structure
where each bit node is only connected to a single check node in one check-vector node,
as shown in Fig~\ref{fig_Equivalent_decoding_tree}. This is essential to obtaining
an enhanced erasure-recovery performance,
as the dependence (cycles) between the constituent bits in $\bar{\mathbf w}$
may introduce more unrecoverable bits with regard to the decoding procedure.

By generalizing the results in example~\ref{eg_inverse_operation}, we give the decoding
algorithm below.
\begin{description}

  \item[Step 1]: For each edge $(n,m)$ connected to variable node $n$
  in $\mathcal G$, we multiply the inverse of its label
  to its associated parity check function in the check node $m$.
  Then, equivalently in $\bar{\mathcal G}$, we get
  the parity check equation set $\sum_{i\neq n}\mathbf A^{k_i-k_n}
  \bar{\mathbf x}_{i}=\bar{\mathbf x}_{n}$ according to Eq.~\eqref{eqn_inverse_operation}
  in check-vector node $\bar{\mathbf m}$ .

  \item[Step 2]: For each constituent check node in
  Fig.~\ref{fig_Equivalent_decoding_tree} connected to a single
  unrecovered bit node in $\bar{\mathbf n}$,
  we recover the value of the erased bit node as the XOR of the other bit nodes participating
  in its parity-check relation.
  \item[Step 3]: Go to Step $1$ until all the bits are recovered or the maximum
  number of iterations is reached.
\end{description}

Actually, the matrix inverse operation is not necessary.
Once the power of the canonical cyclic generator
$\mathbf A$ is determined, its value can be obtained by table-look-up.
The matrix-vector multiplication in Eq.~\eqref{eqn_inverse_operation}
requires a computational complexity $O(\log^2 q)$,
and the processing complexity of the check-vector node relies linearly on the
number of its constituent check nodes, \ie $O(\log q)$. So the overall
check-sum complexity is dominated
by $O(\log^2 q)$.

\begin{figure}[!hbtp]
\begin{center}
\includegraphics[width=0.32\textwidth]{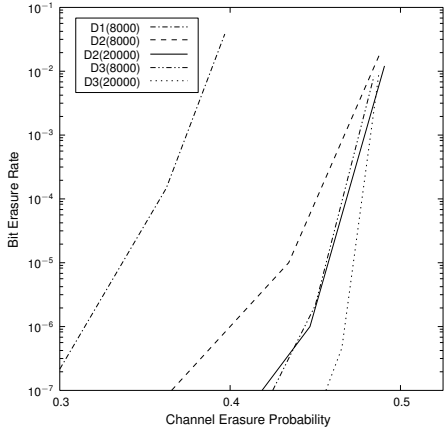}
\end{center}
\vspace*{-.3cm}
\caption{Performance comparison.}
\label{fig_codes_capacities}\vspace*{-.3cm}
\end{figure}

\begin{example}
  Considering the threshold-optimized codes
  \cite{sv08DensityEvolution,ge09NLDPC,Ashi04ExitErasure} of Rate $1/2$
  over $\mathbb F_{4}$, we find the code of threshold-$0.49$ characterized by
  $\lambda(x)=0.72x+0.21x^2+0.06x^4+0.01x^9$ and $\rho(x) = 0.43x^3+0.57x^4$.
  Let $Np$ be the block length. The maximum number of tentative decoding is set to be $60$.
  In order to show that the proposed hybrid iterative decoder can provide an enhanced erasure
  recovery performance, we compare our $4$-ary code with the binary code $D_2(Np)$ characterized by $\lambda(x) = 0.303x + 0.337x^2 + 0.04x^3 + 0.113x^4 + 0.122x^6 + 0.085x^{12}$ and $\rho(x) = 0.85x^5 + 0.15x^6$ with threshold-0.49, and give the performance comparison between
  the decoder $D_3(Np)$ with inverse operation and the decoder $D_1(Np)$ without inverse operation.
  The performance gaps are illustrated in Fig.~\ref{fig_codes_capacities}.
  It can be seen that $D_1$ performs the worst due to the inevitable decoding cycles.
  The hybrid iterative decoder $D_3$ outperforms others as we expect.
\end{example}

Note that in low BER regime the erasure recovery process does not
converge as fast as they do in higher BER regime.
We deal with this problem in section~\ref{sec_convergence_optimize}
and show how to design $\bar{\mathcal C}$ with convergence-optimized degree distribution.


\subsection{Graph Process}
We first introduce a graph process according to \cite{Luby97LossResilient},
which is equivalent to the hybrid iterative decoding process
to facilitate the analysis of the convergence behavior.
Let $\varepsilon_0$ be the initial erasure probability and $\hat{\mathcal G}$
be the graph associated with iterative decoder of $\bar{\mathcal C}$
after putting an inverse operation on a bit-vector node in $\bar{\mathcal G}$.
Note that, $\hat{\mathcal G}$ is a dynamic graph in the sense that every time when we put an inverse operation on an erased or partially erased bit-vector node in $\bar{\mathcal C}$, there is a resulting $\hat{\mathcal G}$. Then, the sub-graph $\hat{\mathcal G}_s$ is defined as the set of all erased bit
nodes together with the check nodes and edges connected to them. The graph process is then described as
follows.
\begin{description}
  \item[Step 1]: On average, there are $\varepsilon_0Np$ bit nodes in the
  initialized graph $\hat{\mathcal G}_s^{(0)}$. We start the graph process at an
  arbitrary check node $\bar{m}^{(0)}$.

  \item[Step 2]: For $\hat{\mathcal G}_s^{(l)}$, $0\leqslant l\leqslant \varepsilon_0Np$,
  if there is only one bit node $\bar{n}^{(l)}$
  connected to $\bar{m}^{(l)}$, we delete the check node $\bar{m}^{(l)}$ and all the check nodes
  with no edge connected to them.

  \item[Step 3]: We delete the bit node $\bar{n}^{(l)}$ together with all its edges. The resulting
  graph is $\hat{\mathcal G}_s^{(l+1)}$.
  \item[Step 4]: Go to Step 2 until all the edges are deleted or maximum number of iterations is reached.
\end{description}
After a \emph{graph operation} (delete) on bit node $\bar{n}^{(l)}$, this bit node is recovered.
Let $\bar{\mathcal N}^{(l)}$ and $\bar{\mathcal M}^{(l)}$ be the set of bit nodes and check nodes in
$\hat{\mathcal G}_s^{(l)}$ respectively, and $\bar{d}_v^{(l)}$ be the
average degree of bit node in $\hat{\mathcal G}_s^{(l)}$.
For any $l$, if the number of check node
$|\bar{\mathcal M}^{(l)}|>|\bar{\mathcal N}^{(l)}|\bar{d}_v^{(l)}/2$ \cite{Luby97LossResilient},
then, there exists a check node $\bar{m}^{(l)}$ that is only
connected to one bit node in $\hat{\mathcal G}_s^{(l)}$.
The graph process will terminate successfully.
Since each graph operation recovers one erased bit node, for sufficiently
large block length $N$ of $\mathcal C$ to recover
$\delta$-fraction of its symbols, the average number of graph operations is
$((1-\delta)^{1/p}-(1-\varepsilon_0))Np$,
which can be steadily obtained by calculating the number of recovered bit nodes.
To achieve a fast convergence performance, the decoder for $\bar{\mathcal C}$ should include as many
graph operations as possible within one iteration,
which is closely related to the degree distribution of ${\mathcal G}$.
In the next section, we show how to design $q$-ary LDPC codes
with convergence-optimized degree distribution.

\subsection{Optimization of the Convergence Performance}\label{sec_convergence_optimize}
The threshold-optimized LDPC code can approach the predicted threshold limit
while the number of decoding iterations tends to be infinity. For finite number of iterations,
a code of non-threshold-optimized distribution may exhibit better convergence property
under some specific channel conditions.
A performance-complexity tradeoff (PCT) for binary LDPC code over Gaussian channel
has been given in \cite{smith10Pct} where they show that the complexity optimization problem
can be reduced to the shaping of the decoding trajectory of EXIT chart for an optimal PCT.
However, the global optimal can not be always guaranteed. Experimental studies show that
the local optimal will suffice \cite{smith10Pct}. One can find more details in \cite{smith10Pct} for
the convex complexity-optimization problem. Since the recursion of the erasure probability can be
used as an EXIT chart to predict the performance threshold,
in the following we adopt the recursion to show that
there also exists
a tradeoff between the convergence rate and the code rate for our hybrid decoder.

We start with estimating the recursion of the erasure probability of our hybrid decoder.
Considering the one-step decoding tree in Fig.~\ref{fig_Equivalent_decoding_tree} for $\hat{\mathcal G}$,
we say a symbol is recovered iff all its constituent bits are recovered.
A constituent bit node $\bar{n}$ can be recovered iff there is a
check node $\bar{m}$ of degree-one connected to it in $\hat{\mathcal G}^{(l)}_s$,
\ie other than the bits in $\bar{\mathbf n}$, there is no erased bits
from the rest bit-vector nodes connected to $\bar{m}$ in $\bar{\mathcal G}$.
Then, it is equivalent to calculate the recursion of bit erasure probability by
\begin{eqnarray}
\label{eqn_bit_era_iter}
    \xi(\varepsilon^{(l)}) = \varepsilon_0\sum_i\hat{\lambda}_i\left(1-
    \sum_j\hat{\rho}_j(1-\varepsilon^{(l)})^{j-d_m}\right)^{i-1},
\end{eqnarray}
where $\varepsilon^{(l)}$ is the bit erasure probability from $l$-th iteration.
$\hat{\lambda}_j$ and $\hat{\rho}_j$ are the degree distribution for $\hat{\mathcal G}$,
which equal to $\bar{\lambda}_j$ and $\bar{\rho}_j$ respectively, and can be calculated by
\begin{equation}
\label{eqn_deg_binary}
  \hat{\rho}_{j} = \frac{\sum_ib_{j,i}j\rho_i/i}{\sum_i\sum_{j=i}^{pi}b_{j,i}j\rho_i/i},
\end{equation}
where $b_{j,i}$ is the probability that a check node within a degree-$i$
check-vector node is connected to $j$ bit nodes which is taken from the polynomial
$f_p(x) = (a_1x+...+a_px^p)^{i}=\sum_{j=i}^{pi}b_{j,i}x^j$,
and $a_i$ is the probability that the row weight of a random matrix label is $i$.
$d_m = \sum_{i=1}^{p} ia_i$ is the average row weight.
If the labels are generated randomly with uniform distribution,
$a_i = {p\choose i}/(q-1)$. $\hat{\lambda}_{j}$ can be calculated in the same way,
\ie replacing $\rho_j$ by $\lambda_j$ in Eq.~\ref{eqn_deg_binary}.
Then, the recursion of symbol erasure probability $\gamma$ is simply
$\gamma^{(l+1)}=1 - (1-\phi(\gamma^{(l)}))^{p}$,
where $\phi(\gamma^{(l)})=\xi(1-(1-\gamma^{(l)})^{1/p})$.

Let $L$ be the number of iterations done in the decoder, $\gamma_L$ be the fraction
of symbols that are not recovered.
Then convergence performance can be characterized by
the average number of graph operations per iteration, which is given by
$g(\gamma_L)=\left((1-\gamma_L)^{(1/p)}-(1-\varepsilon_0)\right)Np/L$.
$L$ can be calculated by
$L = \int_{\gamma_L}^{\gamma_0}{\left( \gamma\ln \left( \frac{\gamma}{f(\gamma)} \right)\right)^{-1}d\gamma}$
\cite{smith10Pct},
where $f(\gamma)=\gamma^{(l+1)}$ and $\gamma_0$ is the initial symbol erasure probability.
Setting $R_0\leqslant R$, we have the optimization algorithm below.
\begin{eqnarray}
\label{eqn_converge_algorithm}
\nonumber
  \text{maximize} && \frac{((1-\gamma_L)^{1/p}-(1-\varepsilon_0))Np}{\int_{\gamma_L}^{\gamma_0}{\left( \gamma\ln \left( \frac{\gamma}{f(\gamma)} \right)\right)^{-1}d\gamma}}.\\
\nonumber
  \text{subject to}
  && \gamma < f(\gamma);\\ \nonumber
  && \sum_i(\lambda_i/i) \geqslant \frac{\sum_i(\rho_i/i)}{1-R_0};\\
\nonumber
  && \lambda_i\geqslant 0,\rho_i \geqslant 0;\\
\nonumber
  && \sum_{i} \lambda_i = \sum_{i} \rho_i = 1;\\
  && \|{\lambda}-\bar{ \lambda}\|_\infty<\zeta_1, \|\rho-\bar{\rho}\|_\infty<\zeta_2,
\end{eqnarray}
where $\bar{\lambda}$ and $\bar{\rho}$ can be initialized as the threshold-optimized LDPC
codes suggest \cite{sv08DensityEvolution,ge09NLDPC,Ashi04ExitErasure}.
$R_0$ is fixed which is lower than the rate of the code $(\bar{\lambda},\bar{\rho})$.
$\zeta_1$ and $\zeta_2$ are carefully set to be small values to guarantee
finding the unique local maximum \cite{smith10Pct}.
The constraint $\gamma < f(\gamma)$ is substantial such that this optimization algorithm is valid.
This \emph{irregular} algorithm is different from the \emph{quasi-regular} algorithm in \cite{smith10Pct}
in the sense that we update $\bar{\lambda}$ and $\bar{\rho}$
by the recent optimal values in each iteration
through which we obtain the convergence-optimized $q$-ary LDPC codes.

\begin{table}
  \caption{The accuracy of the estimated degree distribution.}
  \begin{center}
      \begin{tabular}{||c||c|}
        \hline
        \multirow{3}*{$\hat{\lambda}^{(e)}(x)$} &
        $0.05563x+0.16690x^2+0.19490x^3+0.14002x^4$\\
        &$+0.12261x^5+0.13245x^6+0.10722x^7+0.05676x^8$\\
        &$+0.01927x^9+0.00385x^{10}+0.00035x^{11}$ \\ \hline
        \multirow{3}*{$\hat{\lambda}(x)$} &
        $0.05564x+0.16688x^2+0.19493x^3+0.13998x^4$\\
        &$+0.12260x^5+0.13239x^6+0.10718x^7+0.05684x^8$\\
        &$+0.01918x^9+0.00390x^{10}+0.00044x^{11}$ \\ \hline
      \end{tabular}
   \end{center}
   \label{tb_est_degree}
\end{table}

\begin{figure}[!hbtp]
\begin{center}
\includegraphics[width=0.32\textwidth]{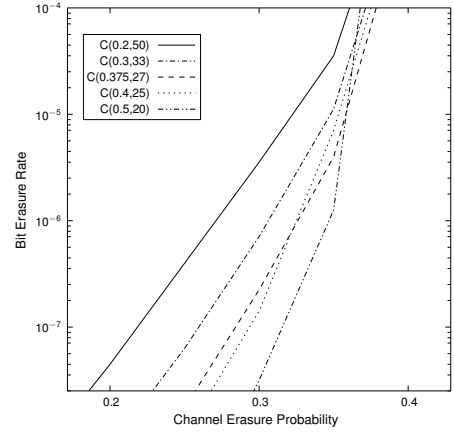}
\end{center}
\vspace*{-.3cm}
\caption{Convergence performance comparison.}
\label{fig_pct}\vspace*{-.3cm}
\end{figure}

\begin{example}\label{eg_optimized_algorithm}
  First, we show
  the accuracy of the estimated degree distribution made by Eq.~\eqref{eqn_deg_binary}.
  We consider the code of length $20000$-bits over $\mathbb F_8$
  whose variable degree distribution is $\lambda(x)=0.5x+0.5x^3$.
  In table~\ref{tb_est_degree}, we give the estimated degree distribution $\hat{\lambda}^{(e)}(x)$
  and the actual degree distribution $\hat{\lambda}(x)$ for the equivalent binary code.
  It can be seen that estimation made by Eq.~\eqref{eqn_deg_binary} are actually very accurate.

  Then, we accelerate the convergence process from bit erasure probability $10^{-3}$ to $10^{-7}$
  for $q$-ary LDPC codes of length $10000$-bits over $\mathbb F_8$.
  Let $L$ be the number of iterations and $C(g(\gamma_L),L)$ be the designed codes.
  We calculate BERs by the average of 100 time experiments
  and compare the erasure-recovery performance in Fig~\ref{fig_pct}.
  $C(0.375,27)$ is the threshold-optimized code with threshold-$0.3998$
  \cite{sv08DensityEvolution,ge09NLDPC,Ashi04ExitErasure}
  characterized by
  $\lambda(x)=0.71x+0.23x^3+0.03x^4+0.01x^7+0.02x^{11}$ and $\rho(x) = 0.32x^4+0.68x^5$.
  Then we obtain the convergence-optimized code $C(0.5,20)$ with threshold-$0.356$
  characterized by
  $\lambda(x)=0.45x+0.18x^2+0.15x^3+0.03x^5+0.08x^{8}+0.11x^{13}$ and $\rho(x) = 0.27x^4+0.73x^5$.
  The optimized code $C(0.5,20)$ outperforms other codes.
  The convergence process is accelerated by $33\%$ with regard to $g(\gamma_L)$
  and $26\%$ with regard to $L$.
  We note that the values of $g(\gamma_L)$ are smaller in the low BER regime,
  which means that only a small fraction of graph operations are included
  in the area during each iteration.
  The optimization of convergence process tries
  to find a tradeoff between the code rate and the convergence rate.
\end{example}

\section{Discussions and Conclusion}
In this letter, we show how to design convergence-optimized $q$-ary LDPC
code over BEC by introducing a hybrid iterative decoder.
Different to other non-binary decoders,
this one can be characterized by using binary analysis tools.
In addition, extra benefit coming from the equivalent binary representation
is that the binary decoder can be easily concatenated to its associated non-binary decoder.

\ifCLASSOPTIONcaptionsoff
  \newpage
\fi

\bibliographystyle{IEEEtran}
\bibliography{IEEEabrv,Chen_WCL2012-0297}

\begin{thebibliography}{1}
\providecommand{\url}[1]{#1}
\csname url@samestyle\endcsname
\providecommand{\newblock}{\relax}
\providecommand{\bibinfo}[2]{#2}
\providecommand{\BIBentrySTDinterwordspacing}{\spaceskip=0pt\relax}
\providecommand{\BIBentryALTinterwordstretchfactor}{4}
\providecommand{\BIBentryALTinterwordspacing}{\spaceskip=\fontdimen2\font plus
\BIBentryALTinterwordstretchfactor\fontdimen3\font minus
  \fontdimen4\font\relax}
\providecommand{\BIBforeignlanguage}[2]{{%
\expandafter\ifx\csname l@#1\endcsname\relax
\typeout{** WARNING: IEEEtran.bst: No hyphenation pattern has been}%
\typeout{** loaded for the language `#1'. Using the pattern for}%
\typeout{** the default language instead.}%
\else
\language=\csname l@#1\endcsname
\fi
#2}}
\providecommand{\BIBdecl}{\relax}
\BIBdecl

\bibitem{sv08DensityEvolution}
\BIBentryALTinterwordspacing
V.~Savin, ``Non binary ldpc codes over the binary erasure channel: density
  evolution analysis,'' \emph{CoRR}, vol. abs/0810.4404, 2008, informal
  publication. [Online]. Available:
  \url{http://dblp.uni-trier.de/db/journals/corr/corr0810.html}
\BIBentrySTDinterwordspacing

\bibitem{ge09NLDPC}
G.~Li, I.~Fair, and W.~Krzymien, ``Density evolution for nonbinary ldpc codes
  under gaussian approximation,'' \emph{IEEE Transactions on Information
  Theory}, vol.~55, no.~3, pp. 997--1015, march 2009.

\bibitem{Ashi04ExitErasure}
A.~E. Ashikhmin, G.~Kramer, and S.~ten Brink, ``Extrinsic information transfer
  functions: model and erasure channel properties,'' \emph{IEEE Transactions on
  Information Theory}, vol.~50, no.~11, pp. 2657--2673, 2004.

\bibitem{Savin09Bec}
V.~Savin, ``Binary linear-time erasure decoding for non-binary ldpc codes,'' in
  \emph{Information Theory Workshop}, oct. 2009, pp. 258--262.

\bibitem{lidl2002}
R.~Lidl, \emph{Introduction to finite fields and their applications.}\hskip 1em
  plus 0.5em minus 0.4em\relax Cambridge, 2002.

\bibitem{xiao07UniDecLdpc}
X.~Ma and B.~Bai, ``A unified decoding algorithm for linear codes based on
  partitioned parity-check matrices,'' in \emph{Information Theory Workshop},
  sept. 2007, pp. 19--23.

\bibitem{Luby97LossResilient}
\BIBentryALTinterwordspacing
M.~Luby, M.~Mitzenmacher, M.~A. Shokrollahi, D.~A. Spielman, and V.~Stemann,
  ``Practical loss-resilient codes.'' in \emph{STOC}, 1997, pp. 150--159.
  [Online]. Available:
  \url{http://dblp.uni-trier.de/db/conf/stoc/stoc1997.html}
\BIBentrySTDinterwordspacing

\bibitem{smith10Pct}
B.~Smith, M.~Ardakani, W.~Yu, and F.~Kschischang, ``Design of irregular ldpc
  codes with optimized performance-complexity tradeoff,'' \emph{IEEE
  Transactions on Communications}, vol.~58, no.~2, pp. 489--499, February 2010.

\end{thebibliography}

\end{document}